\documentclass[epj]{svjour}
\usepackage{graphics}
\usepackage{graphicx}
\begin{document}
\title{Sums of variables at the onset of chaos}
\author{Miguel Angel Fuentes\inst{1} \and Alberto Robledo\inst{2}% etc
}                     % Do not remove
\offprints{}          % Insert a name or remove this line
\institute{Centro de Investigaci\'on en Complejidad Social, Facultad de Gobierno, Universidad del Desarrollo, Santiago, Chile\\
              Santa Fe Institute, 1399 Hyde Park Road, Santa Fe, New Mexico 87501, USA \\
              Centro At\'omico Bariloche, Instituto Balseiro and CONICET, 8400 Bariloche, Argentina\\
              \email{fuentesm@santafe.edu}           %  \\
           \and
              Instituto de F\'{\i}sica  y Centro de Ciencias de la Complejidad, Universidad Nacional Aut\'onoma de M\'exico,
Apartado Postal 20-364, M\'exico 01000 DF, Mexico
}
\date{Received: date / Revised version: date}
\abstract{
We explain how specific dynamical properties give rise to the limit
distribution of sums of deterministic variables at the transition to chaos
via the period-doubling route. We study the sums of successive
positions generated by an ensemble of initial conditions uniformly
distributed in the entire phase space of a unimodal map as represented by
the logistic map. We find that these sums acquire their salient, multiscale,
features from the repellor preimage structure that dominates the dynamics
toward the attractors along the period-doubling cascade. And we explain how
these properties transmit from the sums to their distribution. Specifically, we show
how the stationary distribution of sums of positions at the Feigebaum point
is built up from those associated with the supercycle attractors forming a
hierarchical structure with multifractal and discrete scale invariance
properties.
 % end of PACS codes
} %end of abstract
\maketitle
\section{Introduction}
\label{intro}

Sums of deterministic variables, such as those generated from consecutive
positions of trajectories of iterated one-dimensional nonlinear maps, lead
to limit distributions that reflect the periodic or chaotic character of the
dynamics at work. In the former case when the Lyapunov exponent is negative, 
$\lambda <0$, the distribution of trajectories initiated within the
attractor is trivially determined by its finite set of positions. In the
latter case, when $\lambda >0$,\ a single chaotic trajectory (or an ensemble
of them) leads to a gaussian stationary distribution, in line with the
ordinary central limit theorem, just as independent random variables do \cite{vankampen1} \cite{khinchin1}. The limiting case of vanishing Lyapunov exponent $%
\lambda =0$\ poses interesting questions about the nature of its stationary
distribution: Does this borderline case lead to a general type of stationary
distribution, or, on the contrary, to distributions that capture the
particular features of the dynamics involved? \ Do the absence of ergodicity
and mixing preclude the appearance of broad-spectrum distributions
compatible with statistical-mechanical theories? There has been speculation
and discussion \cite{tsallis1} \cite{grassberger1} about whether sums of
correlated deterministic variables at vanishing, or near vanishing, Lyapunov
exponent give rise to a general type of nongaussian stationary distribution.
The well-known multifractal attractor at the period-doubling transition to
chaos, the Feigenbaum attractor, has proved to be a suitable model system
for the exploration of this issue \cite{tsallis1} \cite{grassberger1}.

We have previously provided an answer \cite{robledo1} \cite{robledo2} for the case
of trajectories initiated within the Feigenbaum attractor and here we
present definite conclusions for the more involved case of the dynamics
towards this attractor. In the former case the support of the stationary
distribution is the multifractal set that makes up the Feigenbaum attractor
and its amplitude follows its multiscaling property. For the latter case we
demonstrate at this time that the stationary distribution possesses an
infinite-level hierarchical structure that originates from the properties of
the repellor set and its preimages. This ladder organization can be more
easily understood by consideration of the family of periodic attractors,
conveniently, the supercycle attractors \cite{schuster1}, along the
period-doubling cascade. We bring to a close the clarification of this issue.

Specifically, we consider the logistic map $f_{\mu }(x)=1-\mu x^{2}$, $%
-1\leq x\leq 1$, $0\leq \mu \leq 2$, for which the control parameter value
for its main period-doubling cascade accumulation point is $\mu =$ $\mu
_{\infty }=1.401155189092..$. The dynamics at and toward the Feigenbaum
attractor is now known in much detail \cite{robledo3} \cite{robledo4} and this
makes it possible to analyze the properties of sums of iterated positions
that advance to this attractor with the same kind of analytic reasoning and
numerical detail. A fundamental property in the analysis is the following:
Time evolution at $\mu _{\infty }$ from $t=0$ up to $t\rightarrow \infty $
traces the period-doubling cascade progression from $%
%TCIMACRO{\U{3bc} }%
%BeginExpansion
\mu
%EndExpansion
=0$ up to $%
%TCIMACRO{\U{3bc} }%
%BeginExpansion
\mu
%EndExpansion
_{\infty }$ \cite{robledo3} \cite{robledo4}. Beyond a close resemblance between
these two developments there is asymptotic quantitative agreement. Thus, the
trajectory inside the Feigenbaum attractor with initial condition $x_{0}=0$,
takes positions $x_{t}$ such that the distances between nearest neighbor
pairs of them reproduce the diameters $d_{n,m}$ \cite{schuster1}\ defined
from the superstable, or supercycle, orbits of period $2^{n}$, $n=1,2,3,...$%
, with $\overline{%
%TCIMACRO{\U{3bc} }%
%BeginExpansion
\mu
%EndExpansion
}_{n}<%
%TCIMACRO{\U{3bc} }%
%BeginExpansion
\mu
%EndExpansion
_{\infty }$. See Fig. 1. This property has been central to obtain rigorous
results for the fluctuating sensitivity to initial conditions $\xi
_{t}(x_{0})$ within the Feigenbaum attractor, as separations at chosen times 
$t$ of pairs of trajectories originating close to $x_{0}$ can be obtained as
diameters $d_{n,m}$ \cite{robledo3}. It has also been essential in
establishing the discrete scale invariance property of the collective rate
of approach of an ensemble of trajectories to this attractor where the total
length of the $d_{n,m}$, $n$ fixed, gives the fraction of trajectories still
away from the attractor at time $t=2^{n}$ \cite{robledo4}.

\begin{figure}[h]
\centering \includegraphics[width=9.0cm,angle=0]{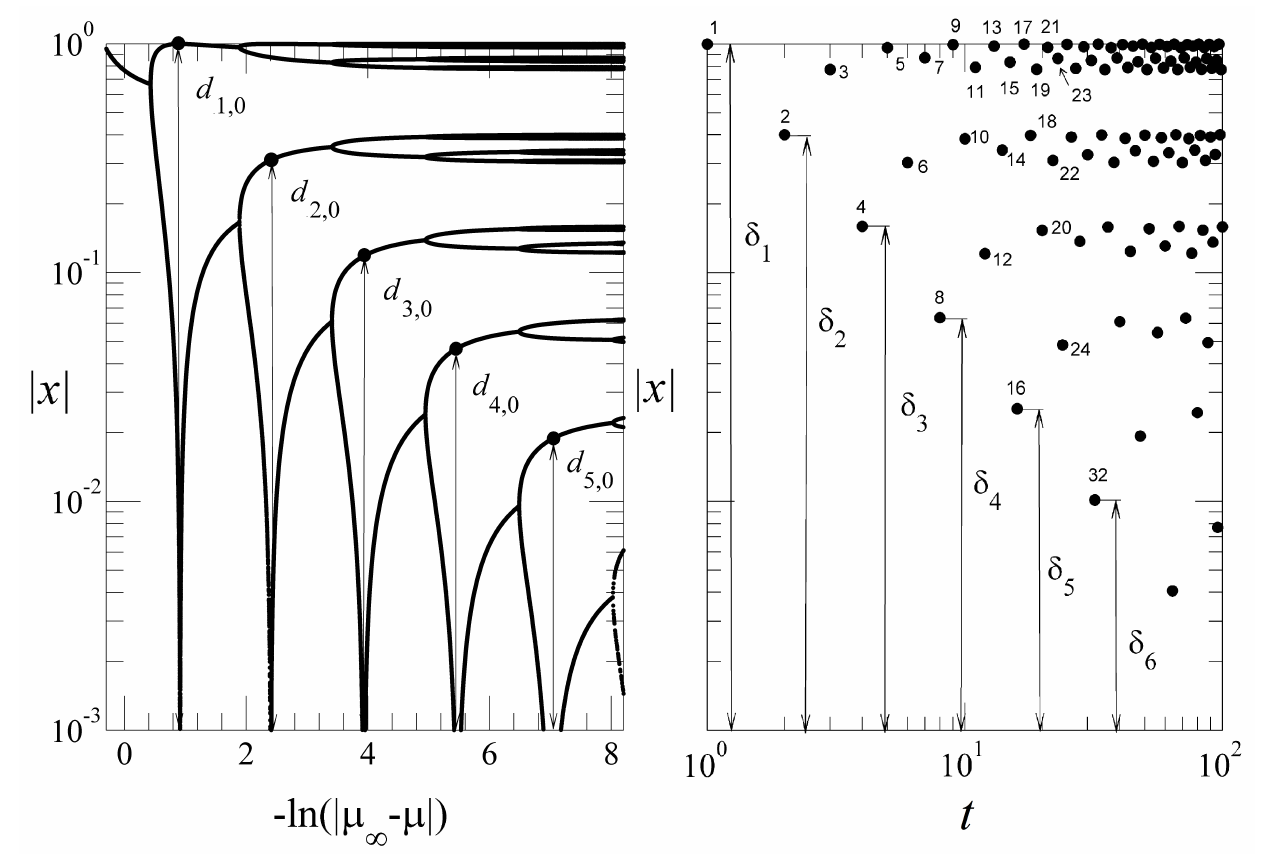}
\caption{{\protect\small Left panel: Absolute value of attractor positions
for the logistic map }${\protect\small f}_{\protect\mu }{\protect\small (x)}$%
{\protect\small \ in logarithmic scale as a function of }${\protect\small -}%
\ln {\protect\small (\protect\mu }_{\infty }{\protect\small -\protect\mu )}$%
{\protect\small . Right panel: Absolute value of trajectory positions for }$%
{\protect\small f}_{\protect\mu }{\protect\small (x)}${\protect\small \ at }$%
{\protect\small \protect\mu }_{\infty }${\protect\small \ with initial
condition }${\protect\small x}_{0}{\protect\small =0}${\protect\small \ in
logarithmic scale as a function of the logarithm of time }${\protect\small t}
${\protect\small , also shown by the numbers close to the circles. The
arrows indicate the equivalence between the diameters }${\protect\small d}%
_{n,0}${\protect\small \ in the left panel, and position differences }$%
{\protect\small D}_{n}${\protect\small \ with respect to }${\protect\small x}%
_{0}{\protect\small =0}${\protect\small \ in the right panel.}}
\label{f1}
\end{figure}

Below we present results for the sums of values of iterated positions
obtained from an ensemble of trajectories initially distributed uniformly
along all the phase space of the logistic map. We evaluated numerically the
sums for $%
%TCIMACRO{\U{3bc} }%
%BeginExpansion
\mu
%EndExpansion
_{\infty }$ and also those for the first few supercycle values $\overline{%
%TCIMACRO{\U{3bc} }%
%BeginExpansion
\mu
%EndExpansion
}_{n}$, $n=1$, $2$ and $3$. Subsequently we determined, numerically also,
the stationary distributions associated with these collections of sums. To
analyze the set of sums of all such positions visited by the ensemble of
trajectories we record their values for all $x_{0}$, $-1\leq x_{0}\leq 1$,
at a fixed large total final time $N$. When $%
%TCIMACRO{\U{3bc} }%
%BeginExpansion
\mu
%EndExpansion
_{\infty }$\ the set of sums is observed to have a complex structure as a
function of $x_{0}$ with very rough, jagged, features that reflect the
properties of the preimage arrangement of positions of trajectories on the
way to the attractor, such as the long journeys taken by initial positions
in the neighborhood of the multifractal repellor. The distribution of these
sums is observed to have an \textit{asymmetric}, \textit{exponentially-decaying},
framework dressed with motifs that involve detail in many scales.

%%%%%%%%%%%%%%%%%%%%%%%%%%%%
%%%%%%%%%%%%%%%%%%%%%%%%%%%%
%%%%%%%%%%%%%%%%%%%%%%%%%%%%
%%%%%%%%%%%%%%%%%%%%%%%%%%%%
%%%%%%%%%%%%%%%%%%%%%%%%%%%%
%%%%%%%%%%%%%%%%%%%%%%%%%%%%

\section{Sums of positions and their distributions for supercycles and their
accumulation point}
\label{sec:1}

We consider the sum of positions $x_{t}$ up to a final iteration time $N$ of
a trajectory with initial condition $x_{0}$ and control parameter value $\mu 
$, \textit{i.e.}

\[
X(x_{0},N;\mu )\equiv \sum\limits_{t=0}^{N}x_{t}. 
\]%
The four panels in Fig. 2 show
the results for $X(x_{0},N;\mu )$ for all possible initial conditions $%
-1\leq x_{0}\leq 1$ and $N\sim O(10^{6})$ when the control parameter takes
the values $\overline{\mu }_{1}$, $\overline{\mu }_{2}$, $\overline{\mu }%
_{3} $ and $\mu _{\infty }$. The plots are all
symmetrical with respect to $x_{0}=0$ and exhibit two large peaks and a
central valley. As it can be clearly observed, $X(x_{0},N;\overline{\mu }%
_{1})$ for period $2$ in panel (a) provides the main frame onto which motifs
of alternating signs are added consecutively to produce $X(x_{0},N;\overline{%
\mu }_{2})$ for period $2^{2}$ in panel (b), and $X(x_{0},N;\overline{\mu }%
_{3})$ for period $2^{3}$ in panel (c). The amplitude of these motifs
decreases rapidly with $n$ such that the fine features in $X(x_{0},N;%
%TCIMACRO{\U{3bc} }%
%BeginExpansion
\mu
%EndExpansion
_{\infty })$ for period $2^{\infty }$ cannot be observed at the scale of the
plot in panel (d). However, zooming on a especially precise evaluation of $X(x_{0},N;%
%TCIMACRO{\U{3bc} }%
%BeginExpansion
\mu
%EndExpansion
_{\infty })$ reveals a delicate and jagged structure that becomes a
multifractal when $N\rightarrow \infty $.

\begin{figure}[tbp]
\centering \includegraphics[width=9.0cm,angle=0]{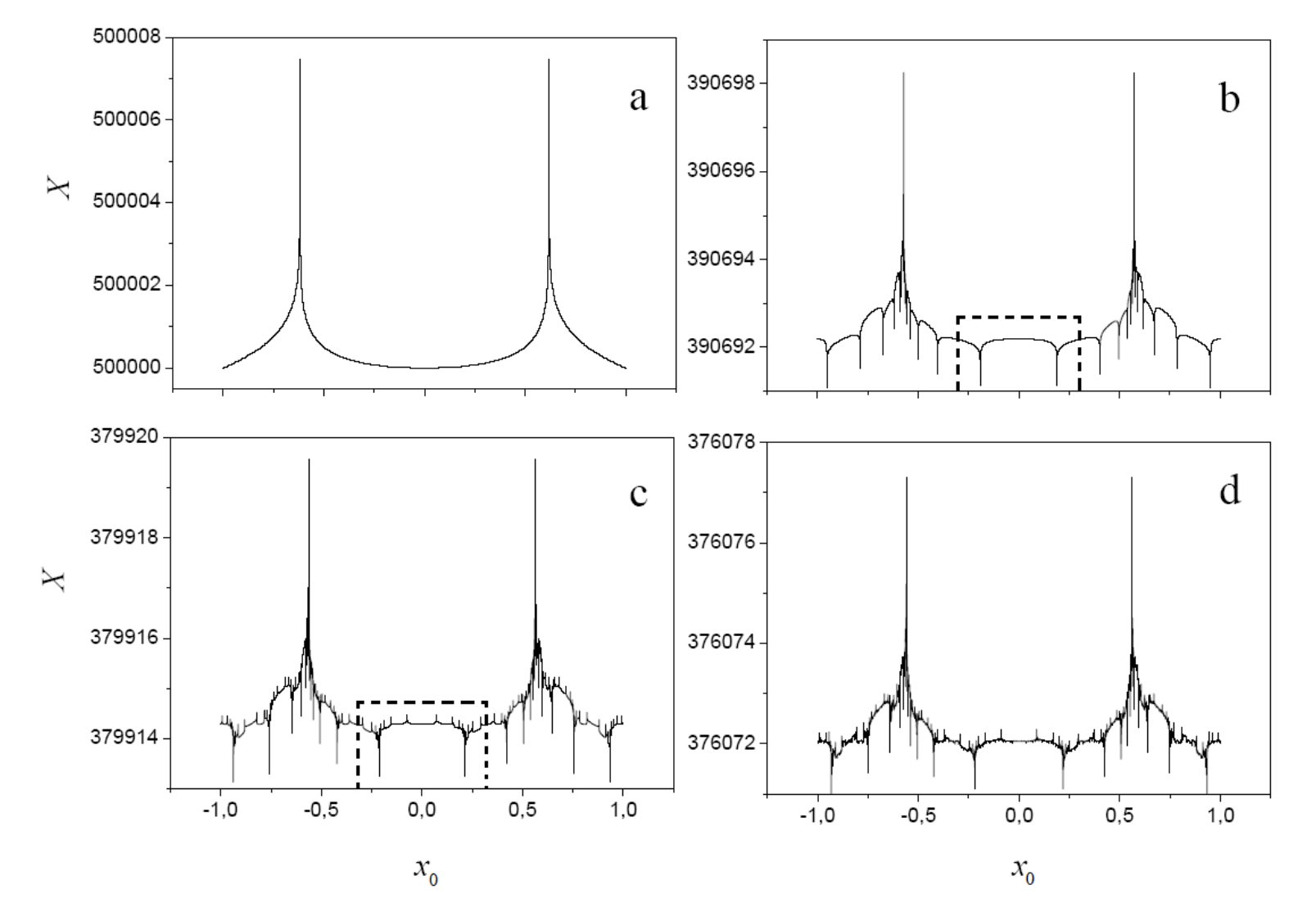}
\caption{Sums $X(x_{0},N;\overline{\mu }_{n})$ as a function of $x_{0}$, $N\sim O(10^{6})$. (a)
Period 2 at $\overline{\mu }_{1}$, (b) period $2^{2}$ at $\overline{\mu }_{2}$, (c) period $2^{3}$ at
$\overline{\mu }_{3}$, and (d) Feigenbaum point $2^{\infty}$ at $\mu _{\infty }$. When inverted and
rescaled the area enclosed within the dashed lines in (b) and (c) become panels (a) and
(b), respectively. See text for details.}
\label{f2}
\end{figure}

To appreciate better the organization of the structures of the sets of sums
we plot in Fig. 3 the right-hand sides with $x_{0}>0$ of the panels in Fig.
2 with a logarithmic scale for the distance between the initial condition $%
x_{0}$ to the position $y$ for the maximum in the sums, i.e. we use the
variables $w_{+}=\ln (x_{0}-y)$ and $w_{-}=\ln (y-x_{0})$. Panel (a) of Fig.
3 shows an almost linear behavior (with slope $\sigma =-1.36$) that indicates
an exponential form around each peak of $X(x_{0},N;\overline{\mu }_{1})$, a
basic element preserved for $\overline{\mu }_{2}$, $\overline{\mu }_{3}$ and 
$\mu _{\infty }$ as can be observed in panels (b), (c) and (d). Next, we
notice that the additional motifs (of smooth protrusions bounded by sharp
drops) in panel (b) for $\overline{\mu }_{2}$ are spaced regularly in the
logarithmic scale, a feature that will let us identify its origin below.
Panel (c) for $\overline{\mu }_{3}$ shows the appearance of additional small
amplitude motifs (this time small smooth valleys bounded by sharp peaks) that
enhance the alternating character to the smooth motifs for $\overline{\mu }%
_{2}$. The scale of panel (d) for $%
%TCIMACRO{\U{3bc} }%
%BeginExpansion
\mu
%EndExpansion
_{\infty }$ precludes the observation of fine features at the transition to
chaos. The insets in panels (c) and (d) help distinguish some differences
between the sums for $\overline{\mu }_{3}$\ and\ $%
%TCIMACRO{\U{3bc} }%
%BeginExpansion
\mu
%EndExpansion
_{\infty }$.

\begin{figure}[tbp]
\centering \includegraphics[width=9.0cm,angle=0]{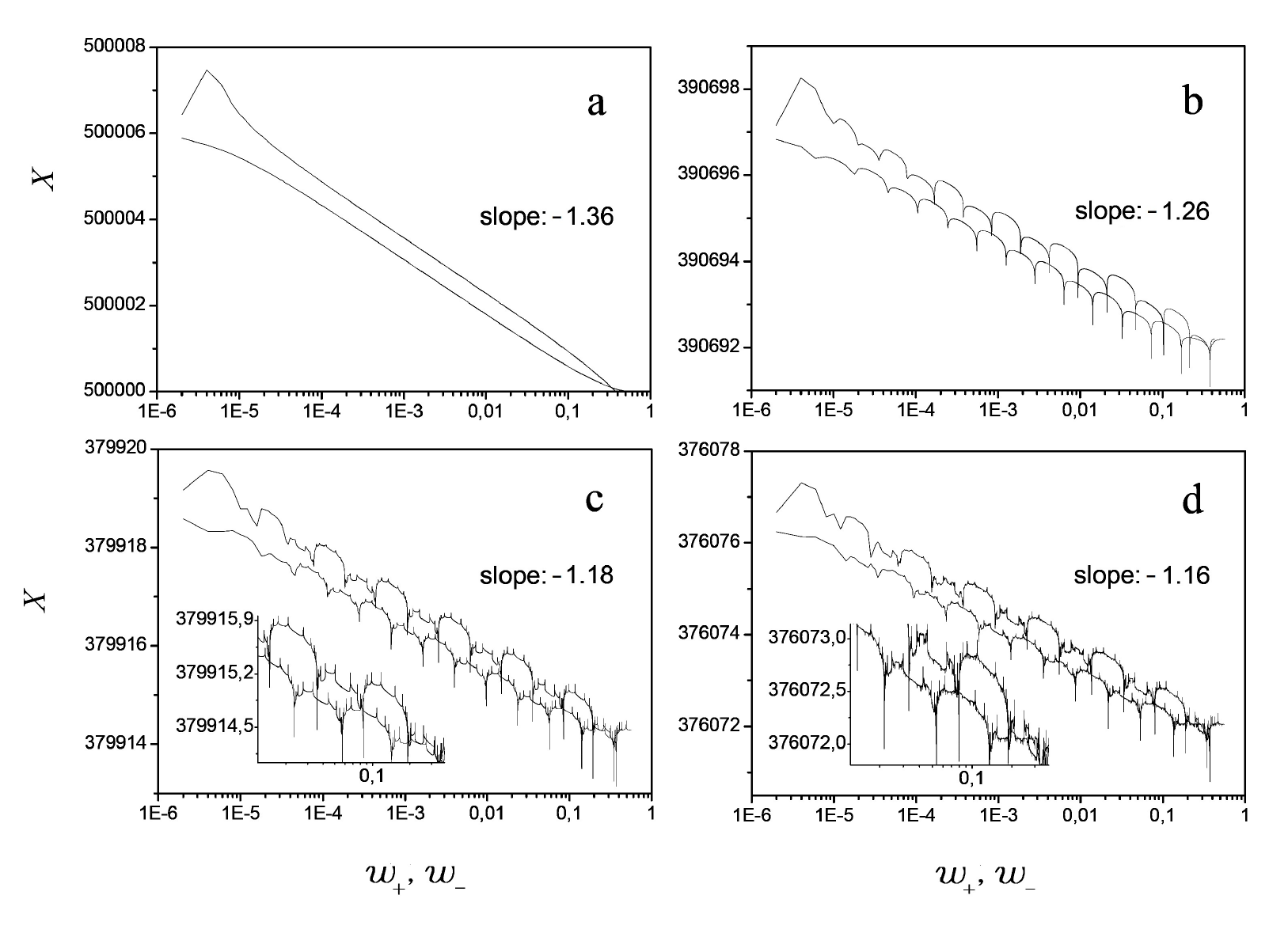}
\caption{Same as Fig. 2 but only the sums with initial conditions $x_{0} >0$ are 
plotted in terms of the new variables $w_{+}=\ln (x_{0}-y)$ (upper curves)
and $w_{-}=\ln (y-x_{0})$ (lower curves). The insets show greater detail.}
\label{f2}
\end{figure}

We proceed to evaluate the distributions associated with the sets of sums $%
X(x_{0},N;\mu )$. The four panels in Fig. 4 show in semi-logarithmic scales
the results for the (non-normalized) histograms that correspond to the sums
in the panels with same labels in Figs. 2 and 3. The essentially linear plot
with slope $\Sigma _{1}=-0.76$ in panel (a) for $\overline{\mu }_{1}$
indicates an anticipated exponential decay distribution. The histogram in
panel (b) for $\overline{\mu }_{2}$\ shows two features in addition to that
in panel (a), a repeated motif structure superposed on the line of slope $%
\Sigma _{1}$ and a new smooth line of slope $\Sigma _{2}=5.02$ that gives the
distribution an overall asymmetrical tent shape (in the semi-logarithmic
scales of Fig. 4). The histogram in panel (c) for $\overline{\mu }_{3}$ adds
to the features of panel (b) both more detail to the motif structure
superposed on the line of slope $\Sigma _{1}=-0.99$ and now a repeated motif structure
superposed on the previously smooth line of slope $\Sigma _{2}=4.87$. These
alternating finer detail appears progressively in the histograms for the
distributions associated with period $2^{n}$ until in the limit $%
n\rightarrow \infty $ we obtain the histogram in panel (d). As before, the
scale of panel (d) for $%
%TCIMACRO{\U{3bc} }%
%BeginExpansion
\mu
%EndExpansion
_{\infty }$ precludes the observation of fine features at the transition to
chaos. The insets in panels (c) and (d) help distinguish some differences
between the histograms for $\overline{\mu }_{3}$\ and\ $%
%TCIMACRO{\U{3bc} }%
%BeginExpansion
\mu
%EndExpansion
_{\infty }$. The stationary distribution for $%
%TCIMACRO{\U{3bc} }%
%BeginExpansion
\mu
%EndExpansion
_{\infty }$ has an asymmetrical double exponential global shape with
superposed motifs of ever-decreasing finer detail characteristic of a
multifractal object. Below we explain the dynamical origin of all the
features of this distribution, but before we rationalize this intricate
arrangement we need to refer to the dynamical properties of trajectories as
they evolve towards the Feigenbaum attractor.

\begin{figure}[tbp]
\centering \includegraphics[width=9.0cm,angle=0]{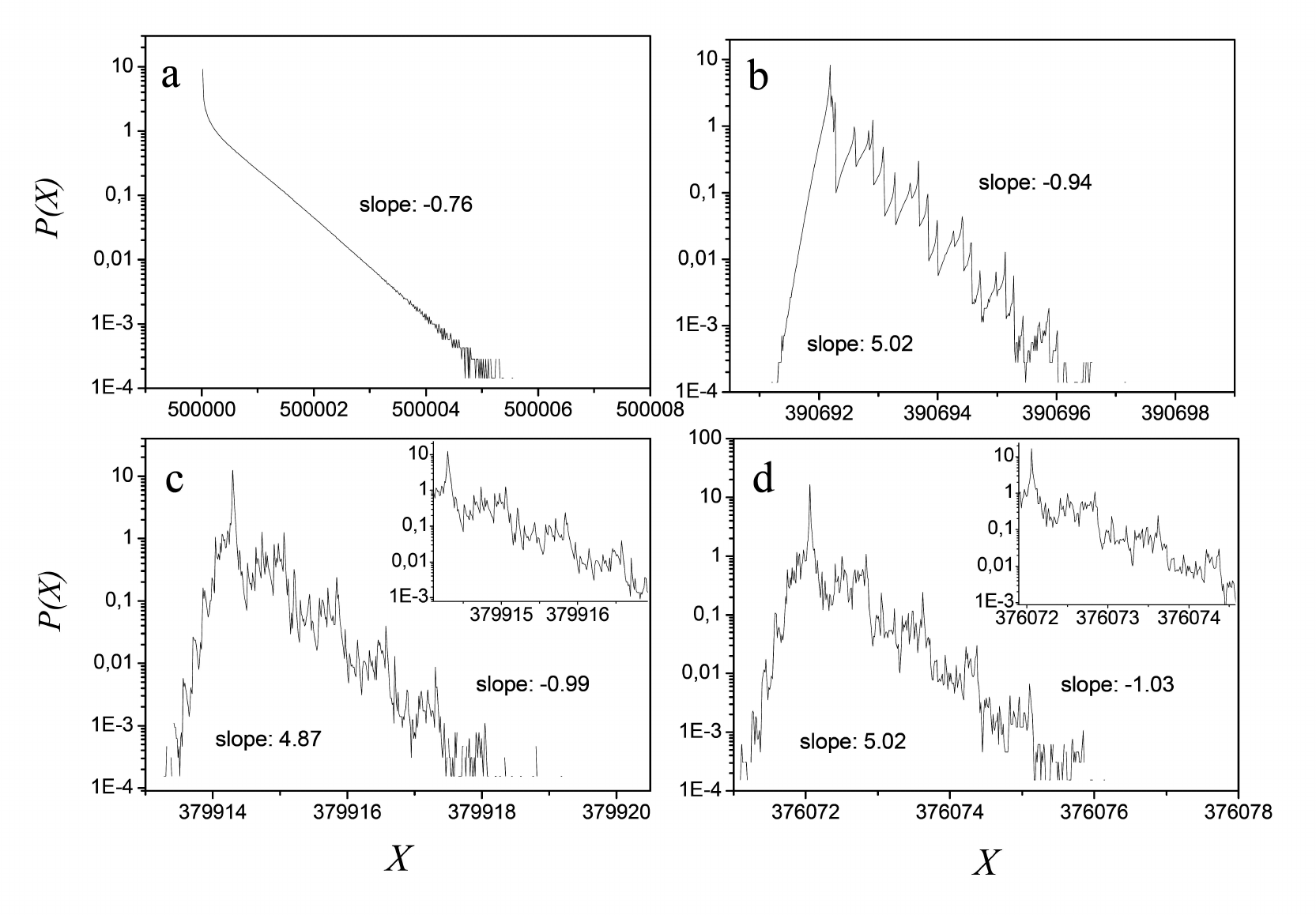}
\caption{Histograms obtained from the sums in Figs. 2 and 3. (a)
Period 2 at $\overline{\mu }_{1}$, (b) period $2^{2}$ at $\overline{\mu }_{2}$, (c) period $2^{3}$ at
$\overline{\mu }_{3}$, and (d) Feigenbaum point $2^{\infty}$ at $\mu _{\infty }$. The insets show greater detail.}
\label{f2}
\end{figure}

%%%%%%%%%%%%%%%%%%%%%%%%%%%%%%
%%%%%%%%%%%%%%%%%%%%%%%%%%%%%%
%%%%%%%%%%%%%%%%%%%%%%%%%%%%%%
\section{Specifics of the dynamics towards supercycles and to their
accumulation point}

For our purposes we recall some basic features of the trajectories at
control parameter values $\mu =\overline{\mu }_{n}$ associated with the
supercycles of periods $2^{n}$, $n=1,2,3,...$, that are located along the
bifurcation forks that form the period-doubling cascade sequence in unimodal
maps (and typically illustrated by the logistic map $f_{\mu }(x)$). The
positions (or phases) of the $2^{n}$-supercycle attractor are given by $%
x_{i}=f_{\overline{\mu }_{n}}^{(i)}(0)$, $i=1,2,\ldots ,2^{n}$. Notice that
infinitely many other sequences of superstable attractors appear at the
period-doubling cascades within the windows of periodic attractors for
values of $\mu >$ $\mu _{\infty }$. Related to the $2^{n}$-attractor at $\mu
=\overline{\mu }_{n}$ there is a $(2^{n}-1)$-repellor consisting of $2^{n}-1$
positions $y_{i}$, $i=1,2,\ldots ,n-1$ (one for $y_{1}$, $2$ for $y_{2}$, $2^{2}$
for $y_{3}$, and so on). These positions are the unstable
solutions, $\left\vert df_{\overline{\mu }_{n}}^{(2^{i})}(y)/dy\right\vert
>1$ , of the equations $y=f_{\overline{\mu }_{n}}^{(2^{i})}(y)$, $%
i=1,2,\ldots ,n-1$. The first, $i=1$, originates at the initial
period-doubling bifurcation, the next two, $i=2$, start at the 2nd
bifurcation, and so on, with the last group of $2^{i}$, $i=n-1$, setting out
from the ($n-1$)-th bifurcation. Other families of periodic attractors along
the period-doubling cascade share most of the properties of supercycles. We
consider explicitly the case of a map with quadratic maximum but the results
are easily extended to general nonlinearity $z>1$.

The organization of the entire set of trajectories generated by all possible
initial conditions (\textit{i.e.} $-1\leq x_{0}\leq 1$) as they flow towards a period 
$2^{n}$-supercycle attractor has been studied in detail \cite{robledo4}. It
was found that the paths taken by the full set of trajectories in their way
to the attractors (or to their complementary repellors) are extraordinarily
structured. To visualize this organization, define the preimage $x^{(k)}$ of
order $k$ of position $x$ to satisfy $x=h^{(k)}(x^{(k)})$ where $h^{(k)}(x)$
is the $k$-th composition of the map $h(x)\equiv f_{\overline{\mu }%
_{n}}^{(2^{n-1})}(x)$. The preimages of the attractor of period $2^{n}$, $%
n=1,2,3,...$ are distributed into different basins of attraction, one for
each of the $2^{n}$ phases (positions) that compose the cycle. That is, a
trajectory initiated at $x^{(k)}$ arrives after $k$ iterations at the attractor, entering
\textit{first} at a given position of the attractor that defines its basin \cite{note1}. When $n>1$
these basins are separated by fractal boundaries whose complexity increases
with increasing $n$. The boundaries consist of the preimages of the
corresponding repellor and their positions cluster around the $2^{n}-1$
repellor positions according to an exponential law. As $n$ increases the
structure of the basin boundaries becomes more involved. Namely, the
boundaries for the $2^{n}$-cycle develops new features around those of the
previous $2^{n-1}$-cycle boundaries, with the outcome that a hierarchical
structure arises, leading to embedded clusters of clusters of boundary
positions, also crowded exponentially, and so forth. The dynamics of
families of trajectories always displays a characteristically concerted
order in which positions are visited, that in turn reflects the repellor
preimage boundary structure of the basins of attraction. That is, each
trajectory has an initial position that is identified as a preimage of a
given order of an attractor (or repellor) position, and this trajectory
necessarily follows the steps of other trajectories with initial conditions
of lower preimage order belonging to a given chain or pathway to the
attractor (or repellor). When the period $2^{n}$ of the cycle increases the
dynamics becomes more involved with increasingly more complex stages that
reflect the hierarchical structure of preimages. See Figs. 2 to 11 in Ref. 
\cite{robledo4} and details therein. The fractal features of the boundaries
between the basins of attraction of the positions of the periodic orbits
develop a structure with hierarchy, and this in turn reflects on the
properties of the trajectories. The set of trajectories produce an ordered
flow towards the attractor or towards the repellor that mirrors the ladder
structure of the sub-basins that constitute the mentioned boundaries.

Another way by which the preimage structure described above manifests in the
dynamics of trajectories moving to the supercycles of periods $2^{n}$ is via
the successive formation of gaps in phase space $[-1,1]$, that in the limit $%
n\rightarrow \infty $ gives rise to the attractor and repellor multifractal
sets. To observe explicitly this process, in Ref. \cite{robledo4} was
considered an ensemble of initial conditions $x_{0}$ distributed uniformly
across $[-1,1]$ and their positions were recorded at subsequent times.
The set of gaps develops in time beginning with the largest one containing
the $i=1$ repellor, then followed by a set of two gaps associated with the $%
i=2$ repellor, next a set of four gaps associated with the $i=3$ repellor,
and so forth. This process stops when the order of the gaps $i$ reaches $n-1$%
. See Figs. 14 to 16 in Ref. \cite{robledo4} and text therein for more
details.

The rate of convergence of an ensemble of trajectories to the attractor and
to the repellor was measured in Ref. \cite{robledo4} by evaluating a single
time-dependent quantity. A partition of phase space was made of $M_{b}$
equally-sized intervals or bins and a uniform distribution of $M_{c}$
initial conditions placed along the interval $[-1,1]$, with $%
r=M_{c}/M_{b}$ the number $r$ of trajectories per bin. The number of bins $%
W_{t}$ that contain trajectories at time $t$ is shown in Figs. 19 in Ref. 
\cite{robledo4} in logarithmic scales for the first five supercycles of
periods $2^{1}$ to $2^{5}$ where the following features were observed: In
all cases $W_{t}$ shows a similar initial nearly-constant plateau, and a
final well-defined exponential decay to zero. In between these two features there are $%
n-1$ oscillations in the logarithmic scales of the figure. The duration of
the overall decay grows approximately proportionally to the period $2^{n}$
of the supercycle. The dynamical mechanism at work behind the features of
the decay rate $W_{t}$ is as follows: every time the period of a supercycle
increases from $2^{n-1}$ to $2^{n}$ by a shift in the control parameter
value from $\overline{\mu }_{n-1}$ to $\overline{\mu }_{n}$ the preimage
structure advances one stage of complication in its hierarchy. Along with
this, and in relation to the time evolution of the ensemble of trajectories,
an additional set of $2^{n}$ smaller phase-space gaps develops and also a
further oscillation takes place in the corresponding rate $W_{t}$ for finite
period attractors. The rate $W_{t}$ for $\mu _{\infty }$ is shown in Fig. 20
in Ref. \cite{robledo4}, the time evolution tracks the period-doubling
cascade progression, and every time $t$ increases from $2^{n-1}$ to $2^{n}$
the flow of trajectories undergoes equivalent passages across stages in the
itinerary through the preimage ladder structure, in the development of
phase-space gaps, and in logarithmic oscillations in $W_{t}$.

Concisely, each doubling of the period introduces additional modules or
building blocks in the hierarchy of the preimage structure, such that the
complexity of these added modules is similar to that of the total period $%
2^{n}$ system. As shown in Ref. \cite{robledo4}, each period doubling adds
also new components in the family of sequentially-formed phase space gaps,
and also increases in one unit the number of undulations in the
log-periodic power-law decay displayed by the fraction $W_{t}$ of ensemble
trajectories still away at a given time $t$ from the attractor (and the
repellor).

\section{Explanation of the structure of the limit distribution for sums of
positions at the Feigenbaum point}

We are in a position now to understand the structure of the sums in Figs. 2
and 3 and their distributions in Fig. 4 in terms of the dynamical properties
of the trajectories that give rise to them. We start by pointing out that
the peak for $x_{0}>0$ in Fig. 2a is precisely located at the position of
the $i=1$ repellor at $\overline{\mu }_{1}$, $y_{1}=-1+\sqrt{1+4\overline{%
\mu }_{1}}/2\overline{\mu }_{1}\simeq 0.6180340...$, the solution of $%
y_{1}=f_{\overline{\mu }_{1}}(y_{1})$, whereas the position of the
accompanying peak for $x_{0}<0$ in the same panel is located at the
repellor's only preimage $x_{1}^{(1)}=-y_{1}$. Actually, the sums at $%
x_{0}=\pm $ $y_{1}$ are infinite when $N$ $\rightarrow \infty $ but our
calculation yields only peaks of finite height since $N<\infty $ while
additionally the values of $\pm $ $y_{1}$ are not exactly reproduced by the
numerical method. Any other value of $x_{0}$ leads to a finite sum when $N$ $%
\rightarrow \infty $ and to smaller values than those of the peaks when $%
N<\infty $. As $x_{0}$ departs gradually from $\pm $ $y_{1}$ the values of
the sums decrease monotonically generating the shape of the plot in the
panel because the number of iterations required to reach the period-$2$
attractor decrease accordingly \cite{note1}. (See Fig. 2 in Ref. \cite{robledo4}).This
decrement in iteration numbers is exponential in $\left\vert
y_{1}-x_{0}\right\vert $ and is captured by the sums as indicated in Fig.
3a. The histogram in Fig. 4a is an exponentially decreasing function of the
sum $X$ because of the exponential shape of the latter, where the lower values of
$X$ are responsible for the larger values of the histogram and vice versa.
This simple shape provides the backbone of all histograms
of the sums $X(x_{0},N;\mu )$ along the period-doubling cascade.

A drastic difference in the dynamics between period $2$ and period $4$ is
that for the latter there is now an infinite number of preimages for the $%
i=1 $ repellor, now located at $y_{1}=f_{\overline{\mu }_{2}}(y_{1})$,
and the same occurs to its mirror 1st preimage located at $x_{1}^{(1)}=-y_{1}$.
The basins of attraction of the $2^{2}$-attractor positions have now fractal
boundaries. The preimages of either $\pm $ $y_{1}$ on these boundaries
cluster exponentially towards $\pm $ $y_{1}$ (as can be seen in Figs. 4 and
5 in Ref. \cite{robledo4}). The two positions of the $i=2$ repellor, the
solutions of $y_{2}=f_{\overline{\mu }_{2}}^{(2)}(y_{2})$, have each
only single preimages located symmetrically at $x_{2}^{(1)}=-y_{2}$. The
(infinite number of) singular cusp shapes that point downwards in Figs. 2b
and 3b also cluster exponentially towards the positions $\pm $ $y_{1}$ and
appear precisely located on the preimages of either $\pm $ $y_{1}$. The sums
with $x_{0}$ located on the preimages of $\pm $ $y_{1}$ are local minima for
the following reasons: First, the trajectories initiated in the close
vicinity of any such locations flow towards the $i=2$ repellor $y_{2}$ (two
positions $y_{2}^{(1)}$ and $y_{2}^{(2)}$) or to its preimage $-y_{2}$ (two
positions $-y_{2}^{(1)}$ and $-y_{2}^{(2)}$) after which they perform a long
quasi period-$2$ cycle around the two positions $y_{2}^{(1)}$ and $%
y_{2}^{(2)}$ before falling into the period-$4$ attractor (see, for example,
Fig. 12 in Ref. \cite{robledo4}). Second, $y_{2}^{(1)}>0$ and $y_{2}^{(2)}<0$%
. Therefore the flow out of any of the preimages of $\pm $ $y_{1}$involves a
large amount of cancellations in the corresponding sums, more than for any
other initial $x_{0}$, and this leads to the sharp dips in Figs. 2b and 3b.
The exponential spacing of the motifs just described in the plots for $%
X(x_{0},N;\overline{\mu }_{2})$ generate the regular seesaw shape in
semi-logarithmic scales of the corresponding histogram in Fig. 4b. The
fluctuating pattern consists of repetition of four oscillations that are
associated with the sinks just described in the sums since there are four
sets of them, two of which are shown in Fig. 3b and the other two are their
mirror images for $x_{0}<0$. The smooth line of positive slope that creates
the tent shape of the histogram originates from the structures of $X(x_{0},N;%
\overline{\mu }_{2})$ around the 2nd generation repellor positions and their
preimages at $\pm y_{2}^{(1)}$ and $\pm y_{2}^{(2)}$ that have a scaled,
inverted, shape of the total sum $X(x_{0},N;\overline{\mu }_{1})$ in Fig. 2a
(see the central region enclosed by dashed lines in Fig. 2b).

The passage from period $4$ to period $8$ introduces another level of
complication to the dynamics towards the attractor. Where there was before
an infinite number of preimages for the $i=1$ repellor clustering
exponentially around it, there is now an infinite number of such clusters,
formed by preimages clustering exponentially around the period $4$ preimages
with shifted locations from $\overline{\mu }_{2}$ to $\overline{\mu }_{3}$.
The clusters themselves cluster around the $i=1$ repellor, now located at $%
y_{1}=f_{\overline{\mu }_{3}}(y_{1})$. And likewise for its 1st
preimage $-y_{1}$. Where there were before only two preimages to the two
positions of the $i=2$ repellor, there are now an infinite number of
preimages clustering exponentially around them, now located at the solutions
of $y_{2}=f_{\overline{\mu }_{3}}^{(2)}(y_{2})$. And likewise for its
two 1st preimages $-y_{2}$. Additionally, there are now four positions of
the $i=3$ repellor, the solutions of $y_{3}=f_{\overline{\mu }%
_{3}}^{(2^{2})}(y_{3})$, that have each only one preimage located
symmetrically at $x_{3}^{(1)}=-y_{3}$. (For more details see Figs. 6 and 7
in Ref. \cite{robledo4}). We observe now in Figs. 2c and 3c the appearance
of new sets of spikes in $X(x_{0},N;\overline{\mu }_{3})$, and the spikes within each set
cluster exponentially around each of the sharp dips that appeared
1st in Figs. 2b and 3b. These new sets of spikes correspond to the clusters
of repellor preimages that, like the clusters themselves, cluster
exponentially around the $i=1$ repellor and its 1st preimage; this because,
as we have seen, the sharp dips do likewise. The exponential spacing of the
clusters of preimages and of the exponential clustering of the preimages
themselves within clusters just described in the plots for $X(x_{0},N;%
\overline{\mu }_{3})$ generate a more intricate seesaw shape in
semi-logarithmic scales of the corresponding histogram in Fig. 4c. The
branch of positive slope that creates the tent shape of the histogram
originates from the structures of $X(x_{0},N;\overline{\mu }_{3})$ around
the 3rd generation repellor positions and their preimages at $\pm
y_{3}^{(1)} $ and $\pm y_{3}^{(2)}$, $\pm y_{3}^{(3)}$ and $\pm y_{3}^{(4)}$%
, that have a scaled, inverted, shape of the total sum $X(x_{0},N;\overline{%
\mu }_{2})$ in Fig. 2b (see the central region enclosed by dashed lines in
Fig. 2c). For this reason we observe that the seesaw pattern complexity of
the positive slope branch of the tent-like histogram for each $2^{n}$%
-supercycle, is equivalent to the negative slope branch of the previous $%
2^{n-1}$-supercycle.

The alternation of upward and downward cusps in the sums $X(x_{0},N;%
\overline{\mu }_{n})$, $n=1,2,3,...$, is due to the fact that the repellor
positions $y_{n-1}$ and 1st preimages $-y_{n-1}$ fall within the bands of
attractor positions shown in Fig. 1b. In this figure (from top to bottom)
the 1st, 3rd, 5th, etc., bands contain all the positive positions, $x>0$,
while the 2nd, 4rd, 6th, etc., bands have all the negative positions, $x<0$.
As partly mentioned, trajectories at $\overline{\mu }_{n}$ initiated close
to $\pm $ $y_{1}$ flow towards the $2^{2}$ positions $\pm y_{2}$, to undergo
a long quasi period-$2$ cycle around the two repellor positions $y_{2}^{(1)}$
and $y_{2}^{(2)}$. Then they proceed to flow towards the $2^{2}$ positions $%
y_{3}$, to undergo a longer quasi period-$2^{2}$ cycle around the $2^{2}$
repellor positions $y_{3}^{(1)}$, $y_{3}^{(2)}$, $y_{3}^{(3)}$, and $%
y_{3}^{(4)}$. This process continues until the $y_{n-1}$ repellor positions
are reached and finally fall into the period-$2^{n}$ attractor (see Fig. 12
in Ref. \cite{robledo4}). These generic flow out of any of the preimages of $%
\pm $ $y_{1}$ (or out of any of the preimages of $\pm $ $y_{i}$, $i=2,3,...n-1$%
) involves a large amount of cancellations in the corresponding sums (more
than for any other $x_{0}$), and leads to sharp cusps. The smallest cusps
point upwards or downwards when $n$ is odd or even, respectively, as this
condition places the $y_{n-1}$ repellor positions within an odd or even band
in Fig. 1b.

Bearing in mind the basic property that trajectories at $\mu _{\infty }$
from $t=0$ up to $t\rightarrow \infty $ trace the period-doubling cascade
progression from $%
%TCIMACRO{\U{3bc} }%
%BeginExpansion
\mu
%EndExpansion
=0$ up to $%
%TCIMACRO{\U{3bc} }%
%BeginExpansion
\mu
%EndExpansion
_{\infty }$, it is clear now how to decode the structure of the sums in
Figs. 2d and 3d and that of their histogram in Fig. 4d. The arrangement of
the multiscale families of cusps of $X(x_{0},N;\mu _{\infty })$ is the
manifestation of the consecutive formation of phase space gaps in an
initially uniform distribution of positions $x_{0}$, and the logarithmic
oscillations at times $t^{k}$, $k=1,2,3,...$, of the rate of convergence of
trajectories $W_{t}$ to the Feigenbaum attractor. The mean time for the
opening a gaps of the same order is $t^{k}$, $k=1,2,3,...$, and the signs
and amplitudes of the cusps are the testimonies stamped in the sums of the
main passages out of the laberynths formed by the preimages of the repellor,
the transits of trajectories from one level of the hierarchy to the next.
Visual representation of the stationary distribution $P(X,N\rightarrow
\infty ;%
%TCIMACRO{\U{3bc} }%
%BeginExpansion
\mu
%EndExpansion
_{\infty })$ associated with the sums $X(x_{0},N;\mu _{\infty })$ is shown
by the histogram in Fig. 4d. It has an asymmetrical double exponential
backbone onto which multiscale patterns are attached that originate from the
aforementioned sets of cusps in the sums. The negative slope (in
semi-logarithmic scales) of the backbone originates from the main $k=1$
repellor core structure shown for all supercycles while the steeper positive
slope originated form the replica structures of the former that originate
from the $k=2$ repellor structures.      

\section{Summary and discussion}

We have shown that the sums of iterated positions of an ensemble of
trajectories moving \textit{toward} the Feigenbaum attractor have a
multiscale, hierarchical, structure that exhibits the preimage organization
of its corresponding repellor. The building blocks of the hierarchy were
identified by looking at the analogous sets of sums obtained from the
dynamics of approach to the simpler supercycle attractors along the
period-doubling cascade. Figs. 2a, 2b and 2c, as well as Figs. 3a, 3b, and
3c, demonstrate clearly how the structure shown in Figs. 2d and 3d for the
sums at the period-doubling accumulation point develops.

This basic property suggests a \ narrow degree of universality for the sums
of deterministic variables at the transitions to chaos, limited to the
universality class of the route to chaos under consideration. Namely, the
sums of positions of memory-retaining trajectories evolving under a
vanishing Lyapunov exponent appear to preserve the particular features of
the multifractal critical attractor and repellor under examination. Thus we
expect that varying the degree of nonlinearity of a unimodal map would
affect the scaling properties of sums or time averages of trajectory
positions at the period-doubling transition to chaos, or alternatively, that
the consideration of a different route to chaos, such as any of the
quasiperiodic routes, would lead to a different structure of comparable time
averages.

As described above, the spiked functional dependence of the sum $X(x_{0},N;%
%TCIMACRO{\U{3bc} }%
%BeginExpansion
\mu
%EndExpansion
_{\infty })$ on $x_{0}$ observable in Figs. 2d and 3d follows the
characteristic hierarchical preimage structure, with exponential clustering,
around the positions of the major and other high-ranking elements of the
repellor and their first preimages \cite{robledo4}. This feature suggests
that large sums of positions are dominated by long journeys toward the
attractor that are particular to the attractor under consideration at the
transition to chaos.

The distributions associated with large sums of positions acquire also a
multiscale, hierarchical, structure, as the nature of $X(x_{0},N;%
%TCIMACRO{\U{3bc} }%
%BeginExpansion
\mu
%EndExpansion
_{\infty })$ is transferred, although in a different setting shown in the
histogram in Fig. 4d, to their distribution $P(X,N;%
%TCIMACRO{\U{3bc} }%
%BeginExpansion
\mu
%EndExpansion
_{\infty })$. Again, the building blocks of the hierarchy in the
distribution are revealed by determination of the analogous distributions of
the sums $X(x_{0},N;\overline{\mu }_{n})$ associated with the supercycle
attractors. Figs. 4a, 4b and 4c show plainly how the structure in Fig. 4d
for the histogram at the period-doubling accumulation point builds up stage
by stage. Parallel to the period-doubling cascade that contains the
supercycles at $\overline{\mu }_{n}<%
%TCIMACRO{\U{3bc} }%
%BeginExpansion
\mu
%EndExpansion
_{\infty }$, there is a chaotic band-splitting cascade at control parameter
values $\widehat{\mu }_{n}>%
%TCIMACRO{\U{3bc} }%
%BeginExpansion
\mu
%EndExpansion
_{\infty }$ that converge also to $%
%TCIMACRO{\U{3bc} }%
%BeginExpansion
\mu
%EndExpansion
_{\infty }$ \cite{schuster1}. The set of chaotic-band attractors at $%
\widehat{\mu }_{n}$ (formed by $2^{n}$ bands) cannot be used as we have done
here for the supercycles at $\overline{\mu }_{n}$ to determine the
stationary distribution $P(X,N;%
%TCIMACRO{\U{3bc} }%
%BeginExpansion
\mu
%EndExpansion
_{\infty })$ because the width of the bands present at $x_{0}$ is gradually
increased as the number of terms in the sums increases and they merge
covering all the interval $[-1,1]$. Because the intraband motion is chaotic
the sums become equivalent to sums of independent random variables and a
stationary gaussian distribution is obtained when $N\rightarrow \infty $ \cite{robledo1} \cite{robledo2}.  

The results presented here are valid for a finite but large number of
summands $N\sim O(10^{6})$, \textit{i.e.} long trajectories, and the numerical
results are dependent also on the fine but finite subdivision of the phase
space interval $[-1,1]$, with a number of bins $M\sim O(10^{6})$ each with a
trajectory initial condition $x_{0}$. Moving towards the limit $M\rightarrow
\infty $ places initial conditions $x_{0}$ closer to the repellor or their
preimage positions enlarging the heights of the spikes in the sums and this
requires increasing $N\rightarrow \infty $ to observe them in Figs. 2 and 3.
On the other hand dumping a number of initial terms in the sum $X(x_{0},N;%
%TCIMACRO{\U{3bc} }%
%BeginExpansion
\mu
%EndExpansion
_{\infty })$ has the effect of clipping the largest spikes in it and leading to
a smoother distribution. However $X(x_{0},N;%
%TCIMACRO{\U{3bc} }%
%BeginExpansion
\mu
%EndExpansion
_{\infty })$ and its distribution are never going to be even, differentiable
functions, for any number of discarded terms. Moreover, the progress toward
a limit distribution is not supposed to involve selected removal of terms.

In the ordinary CLT the sum of independent random variables is equivalent to
the convolution of distributions and as this operation is repeatedly applied
one obtains a gaussian distribution independently of their initial (finite
variance) distribution. The summation process of trajectory positions we
have studied here can be seen as the gradual transformation of an initial
distribution, uniform in our case, into a limiting form by the dynamical
action of the attractor and repellor under vanishing Lyapunov exponent. We
have seen that the particular features of such an attractor/repellor pair
are imprinted in the resulting stationary distribution. \textit{Ad Hoc} attractor/repellor pairs
could be used to construct specific limiting distributions.

%%%%%%%%%%%%%%%%%%%%%%
%%%%%%%%%%%%%%%%%%%%%%
%%%%%%%%%%%%%%%%%%%%%%

\vspace{1cm}

\textbf{Acknowledgements.} MAF thanks CONICYT (Chile): Anillo en Complejidad Social and Proyecto Interfacultades, Universidad del Desarrollo. 
AR acknowledges support from DGAPA-UNAM-IN100311and CONACyT-CB-2011-167978 (Mexican Agencies).

% BibTeX users please use
% \bibliographystyle{}
% \bibliography{}

\end{document}